\documentclass[creativecommons]{eptcs}
\date{}
\usepackage{listings}
\usepackage[inline]{enumitem}
\usepackage{prftree}
\usepackage{amsmath}
\usepackage{amssymb}
\usepackage{graphicx}
\usepackage{hyperref}
\lstdefinestyle{hask}{
      basicstyle=\small\ttfamily,
      flexiblecolumns=false,
      basewidth={0.5em,0.45em},
      literate={+}{{$+$}}1 {/}{{$/$}}1 {*}{{$*$}}1 {=}{{$=$}}1
               {>}{{$>$}}1 {<}{{$<$}}1 {\\}{{$\lambda$}}1
               {\\\\}{{\char`\\\char`\\}}1
               {->}{{$\rightarrow$}}2 {>=}{{$\geq$}}2 {<-}{{$\leftarrow$}}2
               {<=}{{$\leq$}}2 {=>}{{$\Rightarrow$}}2 
               {>>}{{>>}}2 {>>=}{{>>=}}2
               {|}{{$\mid$}}1
               {app}{\$}1
               {/\\}{$\land$}2
               {\\/}{$\lor$}2
               {:iff:}{$\leftrightarrow$}2
               {:neg:}{$\neg$}2
               {:tf:}{$\therefore$}2
               {:seq:}{$\vdash$}2
               {.->.}{$.\rightarrow.$}2
}
\newcommand{\copula}{\mbox{\tt :\$:}}
\title{Carnap: An Open Framework for Formal Reasoning in the Browser}
\author{Graham Leach-Krouse
\institute{Department of Philosophy\\
Kansas State University
Manhattan, KS 66506}
\email{gleachkr@ksu.edu}
}
\def\titlerunning{Carnap: An Open Framework for Formal Reasoning in the Browser}
\def\authorrunning{G. Leach-Krouse}
\begin{document}
\maketitle

\begin{abstract}

  This paper presents an overview of \emph{Carnap}, a free and open framework
  for the development of formal reasoning applications. Carnap's design
  emphasizes flexibility, extensibility, and rapid prototyping. Carnap-based
  applications are written in Haskell, but can be compiled to JavaScript to run
  in standard web browsers. This combination of features makes Carnap ideally
  suited for educational applications, where ease-of-use is crucial for students
  and adaptability to different teaching strategies and classroom needs is
  crucial for instructors. The paper describes Carnap's implementation, along
  with its current and projected pedagogical applications.

\end{abstract}

\section{Introduction}

In what follows we describe the Carnap framework,
a free\footnote{GPLv3-licenced.} and open framework for the development of
formal reasoning applications. One of the goals of the Carnap framework is to
provide instructors with flexible toolkit of interactive online exercises,
widgets, and activities which can introduce students to rigorous formal
reasoning, while minimizing or eliminating pedagogical roadblocks extraneous to
the subject matter.

To that end, Carnap's design emphasizes flexibility and rapid prototyping,
incremental presentation of formal reasoning concepts, and web browser-based
deployment. Carnap's flexibility makes it possible for instructors to adapt
Carnap-powered applications to fit the distinctive needs of their own
institutions, students, and teaching styles. Incremental presentation of
concepts via interactive exercises embedded in instructional materials helps
students focus on what's important rather than being overwhelmed by complexity.
And browser-based delivery of applications (most of Carnap's applications are
compiled from Haskell to JavaScript, allowing them to run in any standard
web browser) removes the early-semester hurdles associated with software
installation, and improves accessibility for technologically anxious students as
well as students who rely on libraries or computer labs for hardware access.

Carnap is under continuing development and is already actively used. In the Fall
semester of 2017, Carnap was the main tool for three different courses at Kansas
State University (Introduction to Formal Logic, Symbolic Logic I, and Modal
Logic). It has also been used supplementally at the University of Birmingham,
and is currently in use at the University of North Carolina-Chapel Hill.

Anecdotal evidence suggests that Carnap is an effective tool for logic
instruction. At Kansas state, student evaluations of classes incorporating
Carnap have consistently been in the top decile of courses university-wide,
which is unusual for a course that is both quantitative and regarded by students
as fairly challenging. Of course, a larger data set and more careful study will
be required before making any definite claims about pedagogical effectiveness. 

However, the following advantages of the framework are clear: a web-based
interface for interactive exercises is highly accessible to students and removes
barriers related to compatibility and installation presented by other forms of
logic-teaching software. And, a flexible core makes it possible to provide for
a wide range of different forms of logic instruction, empowering instructors to
teach in whatever way they feel will be most effective for their students. 

In Sections~\ref{sec:overview_of_the_carnap_framework} and
\ref{sec:theoretical_background} below, we describe the
Carnap framework, together with the theoretical background that has informed
its design.\footnote{Interactive demonstrations of a few applications can be
found at~\cite{carnap}.} We go on, in Section~\ref{sec:applications} to describe
several current educational applications of the framework. We close in Section
\ref{sec:longer_term_prospects} with a discussion of the Carnap framework's
longer-term prospects and projected future applications.

\section{Description of the Carnap Framework}
\label{sec:overview_of_the_carnap_framework}
The Carnap framework consists of three main components. The first of these is
\emph{Carnap-Core}, a set of Haskell libraries which provide:
\begin{enumerate}[label=(\alph*)]
  \item data types for a generic higher-order abstract syntax, henceforth
    \emph{Carnap's syntax}; 
  \item functions applicable to realizations of Carnap's syntax
    that allow for the semantic interpretation, parsing, display, and
    manipulation of all languages based on this syntax;
  \item generic unification algorithms based on the manipulations provided by
    the functions above;
  \item data types for a number of particular languages---in particular,
      languages for propositional logic, propositional modal logic, first-order
      logic, quantified modal logic, monadic second-order logic, and polyadic
      second-order logic---all implemented as realizations of Carnap's abstract
      syntax;
    \item data types for a number of logics for these languages;\footnote{
        By \emph{logic} here, we mean a set of inference rules, rather than a set
        of theorems. So, there are many different propositional logics which can
        prove all and only the arguments whose validity can also be
        demonstrated by truth tables.} and
  \item generic proof-checking methods applicable with these logics.
\end{enumerate}
The second main component of the Carnap framework is \emph{Carnap-GHCJS}, a set of
browser-based educational applications, trans-compiled from Haskell to
JavaScript via the GHCJS compiler~\cite{ghcjs}. These include:
\begin{enumerate}[label=(\alph*)]
  \item an embeddable ``proof-checker'' widget, which uses Carnap's core libraries
    to check the correctness of proofs in any of the logics provided by the
    core, and to provide continuous line-by-line feedback on proofs as they are
    constructed;
  \item a ``rule-builder'' widget which allows students to create and save
    derived inference rules for use in later work;
  \item a JSON API provider, which allows native JavaScript applications to pass 
    JSON proof representations to Carnap for analysis; and
  \item several semantic exercise widgets, including widgets for checking
    student formalizations of natural language statements, and for checking
    student application of semantic methods (e.g.\ truth tables).
\end{enumerate}
The third component of the framework is \emph{Carnap-Server}, a web-server
application based on the Yesod web-development framework~\cite{yesod}. This
application hosts an interactive textbook and instructor-generated assignments
containing interactive exercises, both incorporating the widgets above. It also
manages user accounts for students and instructors, and collects, organizes, and
grades submitted student work. In classes at Kansas state, students generally
work problems by hand in the classroom, where it's possible for the instructor
to provide immediate feedback.\footnote{A few have taken to using Carnap in the
  classroom, on tablets or laptops, for a little extra guidance while working
out problems. This has been permitted.} Outside-of-class assignments are
provided by Carnap-Server, with the Carnap-GHCJS applications supplying rapid
feedback and encouraging students to correct errors before submitting work.

The libraries provided by Carnap-Core have made it possible to rapidly extend
the Carnap Framework to support a wide range of languages and logics. At the
time of writing, Carnap can check proofs in three different styles of natural
deduction---Fitch style proofs, Montegue style proofs, and Hardegree style
proofs---in nine different formal languages, with support for nineteen different
logics, including the propositional and first-order logics of several well-known
open-source logic textbooks and a wide range of relative modal
logics.\footnote{The systems are described in more detail in Hardegree's freely
available online text, Introduction to Modal Logic~\cite{intromodal}.}


\section{Theoretical Background}
\label{sec:theoretical_background}

In this section, we briefly describe the theoretical background that has
informed Carnap's design. In particular, we will detail the design principles
that have motivated several of the choices mentioned above, and will expand upon
the implementation of Carnap's generic higher-order abstract syntax and
unification methods.

\subsection{Design Principles}
\label{sub:design_principles}

The primary motivation for the development of the Carnap framework has been the
desire to provide an educational proof assistant that is simultaneously
\emph{flexible} enough to provide for the needs of a broad curriculum in formal
reasoning and \emph{accessible} enough for use both by students at the very
introductory levels where the majority of logic instruction takes place and by
educators who are not necessarily also software developers.

The demand for flexibility motivates an emphasis on genericity in the basic
components of the framework. If our goal is to cover not just elementary
propositional and first-order logic, but also perennial topics in philosophical
logic, such as modality, plural and higher-order quantification, set theoretic
foundations of mathematics, abstraction-theoretic foundations of mathematics,
formal theories of truth, and so on, we will need to be able to handle a wide
variety of different syntaxes and rules of inference. Writing ad-hoc
proof-checking methods for each of these different logics would be exhausting
and would result in bloated and unmaintainable software. Hence, there is a need
for a notion of syntactical data general enough to encompass the many formal
reasoning systems that Carnap aims to accommodate, and for proof-checking
algorithms generic enough to be applied in combination with any of the sets of
rules of inference that occur in the target systems.

The demand for accessibility motivates the browser-first approach that the
Carnap framework favors for user interactions, rather than the command-line or
dedicated application approaches used by the majority of
proof-assistants.\footnote{One notable exception to this generalization is
  documented in~\cite{cmu}.  However, even this project requires the
  installation of Flash and Java browser plugins, which are increasingly omitted
  from modern browsers because of their long and well-publicized history of
  security problems.
} Even advanced students can have trouble
installing
and configuring new software.  With a large class of beginning students, simply
getting everyone started with a proof-checking program can take days and can be
a constant source of friction throughout the semester. Directing students to
access a webpage, on the other hand, is quite painless. Introducing a proof
assistant through a series of widgets with increasingly extensive functionality
also seems to expose students to much less shock and anxiety than providing them
with a single monolithic interface right out of the box.


Ordinarily these two demands---flexibility and accessibility---and the
approaches adopted to address them---genericity and a browser-first approach to
user interactions---would be in direct tension with one another. The main
programming language currently available for browser-based applications is
JavaScript. But JavaScript is not designed for generic programming or for the
development of large scale applications.\footnote{For example, {\tt import}
  statements were only introduced to the ECMAScript standard as recently as
  2015, and are only now beginning to be implemented natively in browsers
} One possible remedy would be to do the proof-checking work server-side, rather
than client-side. But proof-checking can be computationally demanding, and
genericity often comes at the cost of efficiency. So the server-side solution
will not scale.

Carnap's use of Haskell overcomes this apparent tension. Haskell's type system
includes a number of features that support highly generic programming.  The
Haskell ecosystem also provides excellent support for basic tasks like parsing
and non-deterministic computation that are fundamental to proof-checking.
Recently, several projects have attempted to trans-compile Haskell code to
JavaScript.\footnote{Besides GHCJS, there is also the Fay compiler~\cite{fay}
  and the Haste compiler~\cite{haste}. And besides these there are languages
that attempt to be imitate Haskell's most important features while still
compiling to JavaScript, like PureScript~\cite{purescript} and Elm~\cite{elm}.}
The GHCJS compiler is currently the most ambitious of these. It allows for most
of Haskell's libraries and any language extension supported by the GHC compiler
to be used in writing applications that run in standard web-browsers. By using
Haskell in combination with the GHCJS compiler, Carnap achieves both genericity
and accessibility. And by distributing the computational load of generic
proof-checking across a large number of clients, Carnap's basic model can
smoothly scale to accommodate very large numbers of students without any
noticeable degradation in performance.

\subsection{Syntactic Data}
\label{sub:syntactic_data}

In this section, we explain Carnap's generic syntax, showing how it provides
a framework capable of accommodating a wide variety of formal languages. In the
first part, we informally describe the basics in the abstract, without
requiring any particular familiarity with Haskell or functional programming. In
the next two parts, we describe the implementation of the generic syntax.
Care has been taken to avoid as much technicality as possible in the latter two
parts.  However, some familiarity with the Haskell programming language is
assumed. These more detailed discussions can be skipped without loss of
continuity.

\subsubsection{General Approach}
\label{ssub:general_approach}

An abstract 
syntax of propositional logic might be informally presented in the
following way: 
\begin{quote}
  The language of propositional logic has a lexicon consisting of a countable
  infinity of sentence letters and the five Boolean connectives:
  $\land$,$\lor$,$\rightarrow$, $\leftrightarrow$, and $\neg$.

  The main syntactic category of our language is \emph{formula}. 
  
  A formula is either: a single sentence letter, the result of applying one of
  the binary Boolean connectives $\land$,$\lor$,$\rightarrow$ or
  $\leftrightarrow$ to a pair of formulas, or the result of applying the
  unary connective $\neg$ to a formula
\end{quote}
It is tempting to analyze the above as an informal specification of a simple
recursive data type, with each lexical item a constructor for that type.
However, consider that an informal presentation of pure (monadic) first-order
logic might then proceed to specify:
\begin{quote}
  The language of pure first-order logic extends the lexicon of propositional
  logic with a countable infinity of predicate letters, the quantifiers
  $\forall$ and $\exists$, and a countable infinity of term variables.

  The syntactic categories of first-order logic are \emph{formula} and
  \emph{term}
  
  All term variables are terms. Formulas are built as in propositional logic,
  and also by applying a predicate to a term, or applying a quantifier to bind
  a variable in a given formula.
\end{quote}
Putting this second definition alongside the first calls our attention to
a problem with the tempting analysis above. When, in the previous analysis, we
took the language of propositional logic for a recursive data type and took,
e.g. the connective $\neg$ for a constructor, we committed ourselves to thinking
of $\neg$ as a function taking a formula of propositional logic as input and
returning a formula of propositional logic. But in the language of first-order
logic, at least informally, we find that same connective being applied, now to
formulas of \emph{first-order} logic, and now returning formulas of first-order
logic.

Of course, we could hold on to our original analysis and deny that the $\neg$ of
propositional logic is the same constructor as the $\neg$ of first-order logic,
regarding talk about ``extending the lexicon'' of propositional logic as a mere
fa\c{c}on de parler. But this will lead to a peculiar case of conceptual
double-vision, as we're constantly forced to distinguish the two almost
indistinguishable negations.  And, in the context of designing a computer
program, we'll have an ugly duplication of code that will only worsen as we
consider other languages containing negation.

Carnap's generic syntax avoids this problem by rejecting the analysis of $\neg$
as the constructor of a recursive data type and, correlatively, rejecting the
analysis of languages as simple recursive data types. Instead, Carnap attempts
to mirror the informal practice (suggested by the language specifications above)
of treating different connectives and lexical categories as portable across
languages.  Carnap does this by treating a connective like $\neg$ as the inhabitant
of a non-recursive type\footnote{Technically, not a concrete type, but
a higher-order type-constructor.} of Boolean connectives. Features of the
connective that are invariant across languages (it applies to elements of the
syntactic category ``formula''; it is rendered prefix as ``$\neg$''; in
first-order semantics it reverses truth value\ldots etc.) are encoded at the
level of this type. Other lexical items are treated similarly---each lexical
category (categories like \emph{Boolean connective} and \emph{sentence letter})
can be thought of as a ``module'', to be freely combined with other modules.
Languages are produced by ``snapping together'' modules\footnote{Essentially,
by taking a direct sum of type-constructors.} and then transforming the
resulting collection of non-recursive data-types into a single recursive
data-type using a fixed-point construction (detailed in
Section~\ref{ssub:the_fundamental_data_types}).

Compared to the analysis of a language as a recursive data type, Carnap's
modular approach is somewhat more cumbersome when dealing with just a single
language. But the Carnap framework is intended to support \emph{multiple}
languages. In this situation, the modular approach greatly reduces duplication
of code.  Instead of specifying what each language is, we specify what a variety
of language modules are, and then specify a language by simply stipulating which
modules are ``snapped together'' to make that language.  Algorithms for working
on languages can also be designed generically, either at the level of individual
modules (for example, a conjunctive-normal-form algorithm that can be applied to
any language containing Boolean connectives), or even at the level of the
structure that all languages have in common in virtue of the way that they are
constructed from modules (for example, a unification algorithm that can be
applied to any language---see Section \ref{sub:unification_algorithms}).

\subsubsection{Two examples}
\label{ssub:two_examples}

To flesh out the general remarks of the previous section, we present two
illustrations of how Carnap represents formal languages: the language of
propositional logic, and the language of first order logic.

Carnap's representation of a propositional language like the one specified in
the previous section would look like this: 
\begin{lstlisting}
data Formula

data Connective lang type where
  And :: Connective lang (Formula -> Formula -> Formula)
  Or  :: Connective lang (Formula -> Formula -> Formula)
  If  :: Connective lang (Formula -> Formula -> Formula)
  Iff :: Connective lang (Formula -> Formula -> Formula)
  Neg :: Connective lang (Formula -> Formula)

data Sentence lang type where
  Sentence :: Int -> Sentence lang Formula

type PropositionalLexicon = Sentence :|: Connective

type PropositionalLanguage = FixLang PropositionalLexicon
\end{lstlisting}
We can see a rough correspondence between the informal presentation just given
and the code here. We have two sorts of lexical items (sentences and
connectives) in the informal definition. In the code, each of those two is given
a data type ({\tt Sentence lang type}, and {\tt Connective lang type}
respectively). We had one basic syntactic category---Formula---in the informal
presentation, which corresponds here to the data type {\tt
  Formula}.\footnote{Notice that this type has no constructors. It's
a ``phantom'' type, used exclusively for organizing other types} The syntactic
type of each lexical item is visible in the second argument of its Haskell type.
So the syntactic type of a sentence is {\tt Formula}, while the syntactic type
of a binary connective, which takes two formulas to make a formula, is {\tt
Formula $\rightarrow$ Formula $\rightarrow$ Formula}. The two lexical data types
are combined into a \emph{lexicon} using the {\tt :|:} operator. The language is
then produced by applying the {\tt FixLang} operator to the lexicon, which
performs the fixed-point construction that turns the non-recursive lexicon type
into a language with recursive formation rules.

We'll save investigation of {\tt Fixlang} and the {\tt :|:} operator for the
next section. For now, we'll instead focus on how treating each sort of lexical
item as a separate ``module'' and treating a language as a sum of different
modules allows us to streamline our code and avoid repetition and boilerplate.

The informal presentation of the language of first-order logic described it as
extending the lexicon of propositional logic with a few new constructors.
Carnap's representation can have first-order logic be exactly that:
\begin{lstlisting}
data Term

data Predicate lang type where
  Pred :: Int -> Predicate lang (Term -> Formula)

data Quantifier lang type where
  All :: Quantifier ((Term -> Formula) -> Formula)
  Some :: Quantifier ((Term -> Formula) -> Formula)

data Variable lang type where
  Var :: Int -> Variable lang Term

type FirstOrderLexicon =  PropositionalLexicon 
                      :|: Predicate 
                      :|: Quantifier
                      :|: Variable

type FirstOrderLanguage = FixLang FirstOrderLexicon
\end{lstlisting}
The fact that the syntactic categories of the language of first-order logic
(namely {\tt Term} and {\tt Formula}) extend those of our propositional language
allows Carnap to automatically infer how the Boolean connectives that make up
part of {\tt PropositionalLexicon} extend their domain when they are lifted to
{\tt FirstOrderLexicon}. Hence, we avoid the repetitive boilerplate involved in
respecifying the propositional parts of the language. Furthermore, functions
defined on {\tt Connective} and lifted from there to {\tt PropositionalLanguage}
can now be lifted to {\tt FirstOrderLanguage} automatically. We also inherit,
for both languages, all functions that have been defined to apply to the
inhabitants of arbitrary fix-points {\tt FixLang a}.

\subsubsection{The fundamental data types}
\label{ssub:the_fundamental_data_types}
To see how what was described in the previous section is possible, we'll need to
do a bit more to dissect {\tt :|:} and {\tt Fixlang}. This section will contain
a certain amount of unavoidable technicality, for which we beg the reader's
patience.

The three fundamental data types (simplified for the sake of readability---in
particular, we omit a lambda-abstraction operator from {\tt Copula} that adds some distracting
complexity) supporting Carnap's generic syntax are as follows:\footnote{\
  Carnap's generic syntax is based on an idea roughly analogous to the one
  presented in~\cite{alacarte}, but adds to this idea a phantom type which is used
  for keeping track of the syntactic category of the constructor.
}
\begin{lstlisting}
data Copula lang t where
    (:app:) :: lang (t -> t') -> lang t -> Copula lang t'

data (:|:) f g lang t where
    FLeft :: f lang t -> (f :|: g) lang t
    FRight :: g lang t -> (f :|: g) lang t

data Fix f t where
    Fx :: f (Fix f) t -> Fix f t
\end{lstlisting}

Suppose we are given the language {\tt PropositionalLanguage} described in the
previous section. (Henceforth, for brevity, we'll refer to this language as {\tt
Lang}, we will refer to to the corresponding lexicon {\tt PropositionalLex} as
{\tt Lex}, and we will refer to to the syntactic type {\tt Formula} as {\tt
Form}.) Individual lexical items of {\tt Lang} are of type {\tt Lang $a$}, where
$a$ is determined according to how these lexical items combine syntactically.
Suppose we are given two lexical items from {\tt Lang}: a sentence letter {\tt
P} of type {\tt Lang Form}, and a negation sign $\neg$ of type {\tt Lang (Form
$\rightarrow$ Form)}. The constructor \copula\ for the type {\tt Copula Lang
Form} can then be applied thus: {$\tt \neg\copula P$}, yielding a value of type
{\tt Copula Lang Form}.  Visually:
\begin{displaymath}
  \resizebox{\textwidth}{!}{
  \prftree
    {\prftree
      {\tt \neg :: Lang\ (Form \rightarrow Form)\hspace{10px}}
      {\tt (\copula):: lang\ (t \rightarrow t') \rightarrow lang\ t\rightarrow Copula\ lang\ t'}
        {\tt (\copula) \neg :: Lang\ Form \rightarrow Copula\ Lang\ (Form\rightarrow Form)}
    }
    {\hspace{-30pt}\prftree[noline]{\tt P :: Lang\ Form}}
  {\tt \neg \copula P :: Copula\ Lang\ Form }}
  \end{displaymath}
Of course, one would expect that applying negation to an atomic formula should
yield something of the same type as the original formula. The trick here is as
follows. Having constructed {$\tt \neg\copula P$}, we then apply {\tt FLeft} to
it, yielding a value of type {\tt (Copula :|: g) Lang Form}. And, recall from
the definitions above, {\tt Lang} is in fact an alias for the {\tt FixLang Lex},
which is again an alias for the type {\tt Fix (Copula :|: Lex)}. Hence, we
can take our value of type {\tt (Copula :|: Lex) Lang Form}, and apply the
constructor ${\tt Fx}$, yielding the desired value of type {\tt Lang Form}.
Visually (with dotted lines representing the rewriting of aliases):
\begin{displaymath}
  \resizebox{\textwidth}{!}{
    \prftree[dotted]
        {\prftree
          {\prftree[dotted]
            {\prftree
              {\tt \neg \copula P :: Copula\ Lang\ Form}
              {\hspace{10px}\tt FLeft :: f\ lang\ t \rightarrow (f :|: g)\ lang\ t}
              {\tt FLeft (\neg \copula P) :: (Copula :|: g)\ Lang\ Form }
            }
            {\tt FLeft (\neg \copula P) :: (Copula :|: g)\ (Fix\ (Copula :|: Lex))\ Form }
          }
          {\prftree[noline]{\tt Fx :: f\ (Fix\ f)\ t \rightarrow Fix\ f\ t }
          }
          {\tt Fx (FLeft (\neg \copula P)) :: Fix\ (Copula :|: Lex)\ Form }
        }
        {\tt Fx (FLeft (\neg \copula P)) :: Lang\ Form}
  }
\end{displaymath}
This does seem like rather a lot of circumlocution. But Haskell's typechecking
is static, so the majority of this work is done at compile time, and has no
noticeable effect on performance.

We see now how languages can be put together. There's a contribution both from
a universal component, {\tt FixLang}, which ensures that all languages contain
a copula for applying different lexical items to one another, and from a 
particularizing component: the lexicon specific to the language in question---in
the case above, the propositional lexicon {\tt Lex}.

This shows us how generic programming can be done over all languages constructed
in Carnap. Since all lexical items have types that unify with {\tt FixLang lex a},
generic functions acting on this type can be applied to all lexical items.
Generic functions like this can be derived using Haskell's typeclass mechanisms.
For example, a {\tt show} function, rendering a lexical item as a string, could
be derived generically for our hypothetical language using a little bit of
boilerplate (written only once for all langauges) like this:
\begin{lstlisting}
class Schematize f where
  schematize :: f a -> [String] -> String

instance Schematize (Fix lex) => Show (Fix lex a) where
  show x = schematize x []

instance Schematize (lex (Fix lex)) => Show (Fix lex) where
  schematize (Fx x) = schematize x

instance (Schematize (lex1 lang), Schematize (lex1 lang)) => 
  Schematize ((lex1 :|: lex2) lang a) where
  
  schematize (FLeft x) = schematize x
  schematize (FRight x) = schematize x

instance Schematize lang => Schematize (Copula lang) where
  schematize (x :app: y) xs = schematize x (schematize y [] : xs)
\end{lstlisting}
together with particular Schematize instances for particular lexical categories:

\begin{lstlisting}
instance Schematize (Connective lang) where
  schematize(And) = \(x:y:xs) -> x ++ "/\" ++ y
  schematize(If)  = \(x:y:xs) -> x ++ "->" ++ y
  schematize(Or)  = \(x:y:xs) -> x ++ "\/" ++ y
  schematize(Iff) = \(x:y:xs) -> x ++ ":iff:" ++ y
  schematize(Neg) = \(x:y:xs) -> ":neg:" ++ x

instance Schematize (Sentence lang) where
  schematize(Sentence n) = \_ -> "P_" ++ show n
\end{lstlisting}

The key observation here is that now we can declare how to print different
connectives once and for all, at the level of lexical modules---languages
will then automatically inherit this information on the basis of the modules
that make them up.

A great deal of functionality can be automatically derived in this way,
including not just rendering of formulas, but also parsing, functions for
semantic evaluation and, most importantly from a proof-checking perspective,
algorithms for unification.

\subsection{Unification Algorithms}
\label{sub:unification_algorithms}

From the perspective of proof-checking, the most significant piece of generic
functionality derived for Carnap's languages is support for higher-order
unification.\footnote{Haskell has  a standard library for first-order
unification---{\tt unification-fd}---but not for higher-order unification.} In
this section, we explain how Carnap applies higher-order unification in order to
allow for the declarative specification of logics. As in the previous section,
a small amount of familiarity with Haskell will be assumed. Familiarity with
basic logic and the simply typed $\lambda$-calculus will also be helpful. As
above, however, this section can be skipped without loss of continuity.

The generic deriving mechanism described in the previous section suffices to
provide a wide range of basic syntactic operations---including substitution,
application, abstraction, $\eta$-expansion and $\beta$-normal\-ization---for any
language defined using {\tt FixLang}. Carnap incorporates an implementation of
Huet's algorithm as given by Dowek in~\cite{handbook}. The algorithm is written
exclusively in terms of the basic syntactic operations that are generically
provided for any language defined using {\tt FixLang}.

Huet's algorithm is a (constructive) semi-decision procedure for
\emph{higher-order unification problems}, i.e. problems of the following form:
\begin{quote} Given a finite set of equations $t_i = u_i$ between terms
  of the simply-typed $\lambda$-calculus, does there exist a substitution
  $\theta$, assigning terms to schematic variables, such that for each $i$, the
  results of applying $\theta$ to $t_i,u_i$ have the same normal form?
\end{quote}
The naive presentation of this algorithm is as a non-deterministic type-directed
stateful computation.  The non-determinism and stateful aspects of the algorithm
can be faithfully reproduced as a pure computation in Haskell using
a combination of the {\tt LogicT} monad transformer~\cite{logict} and the {\tt
State} monad. The type-directedness of the algorithm can be reproduced by using
the syntactic types\footnote{E.g. {\tt Formula} for sentence letters, and {\tt
Formula $\rightarrow$ Formula} for unary connectives.} described in the previous
section, along with Haskell's {\tt Data.Typeable} library.  Huet's algorithm is
the best possible, in the sense that the general problem of higher-order
unification is only semi-decidable. However, the cases we deal with in proof
checking (pattern-matching unification, with fairly simple terms) are in fact
decidable, and in practice Huet's algorithm solves them quickly.

How is this applied to proof checking? Consider the naive classroom presentation
of a common pair of inference rules, modus ponens and universal instantiation:
\begin{displaymath}
  \prftree[l]{(MP)}{\phi\rightarrow\psi}{\phi}{\psi}
  \qquad
  \prftree[l]{(UI)}{\forall x \phi(x)}{\phi(\tau)}
\end{displaymath}
Verifying that a particular inference instantiates the rule MP is a matter of
finding a solution to one of a pair of unification problems determined by the
inference. For example, given the inference
\begin{displaymath}
  \prftree{P\land R\rightarrow Q\lor S}{P\land R}{Q\lor S}
\end{displaymath}
We have the unification problems:
\begin{align*}
  P\land R\rightarrow Q\lor S &= \phi\rightarrow\psi& P\land R\rightarrow Q\lor S &= \phi \\ 
  P\land R &= \phi& P\land R &= \phi\rightarrow \psi \\
  Q\lor S &= \psi &Q\lor S &= \psi
\end{align*}
If we view $\phi,\psi$ as schematic variables and $P,Q,R,S, \rightarrow$ as
constants\footnote{
  Some of which, namely $P,Q,R,S$ are essentially of the type
  {\tt Formula}, and others, namely $\rightarrow$, are of higher types, in this
  case {\tt Formula$\rightarrow$Formula$\rightarrow$Formula}.
} then the inference is an instance of the rule MP, so long as one of these
unification problems has a solution---as the first obviously does, namely:
\begin{align*}
  \phi&\leadsto P\land R & \psi &\leadsto Q\lor S 
\end{align*}
Of course, this case required only first-order unification. We only encounter
higher-order terms once we begin verifying inferences that are supported by
essentially first-order inference rules. For example, the inference from
$\forall x (R(x)\land Q(x))$ to $R(f(a))\land Q(f(a))$ is an instance of UI just
in case the unification problem
\begin{align*}
  \forall x (R(x)\land Q(x)) &= \forall x \phi(x)\\
  R(f(a))\land Q(f(a)) &= \phi(\tau)
\end{align*}
has a solution, when $\phi,\tau$ are regarded as schematic variables,
$\forall,R,f,a$ as constants. Which it does:
\begin{align*}
  \phi&\leadsto \lambda x(R(x)\land Q(x)) & \tau & \leadsto f(a) 
\end{align*}
since $\lambda x(R(x)\land Q(x))(f(a))$ normalizes to $R(f(a))\land Q(f(a))$.

So, the problem of checking whether inferences are instances of rules reduces
to the problem of checking whether certain simple higher-order unification
problems have solutions. As a result, it can be solved in complete generality by
an application of Huet's algorithm.

Against the background of a generic implementation of higher-order unification,
developing a checker for a new logic mostly amounts to simply declaring the
inference rules that the logic contains. Carnap approaches this problem by
extending the language of the logic with lexical modules for one or more lexical
categories of schematic variables. So the lexicon of propositional logic is
extended to
\begin{lstlisting}
type PropositionalLexicon = Sentence :|: Connective :|: SchematicSentence
\end{lstlisting}
In the actual implementation, we also keep track of a context for each proof, in
order to handle assumptions that are later discharged, e.g.\ in conditional
proofs. However, besides the addition of a schematic variable for the assumptions
currently in scope, rules can be declared in Carnap in a way that more or less
mirrors how you might write them on the board in a logic classroom. For example,
here are the two rules from above (MP and UI)\footnote{We use Unicode function
  symbols to emphasize the similarity between Carnap's code for a rule and the
  informal blackboard presentation that that rule might receive.}
\begin{lstlisting}
modusPonens = [ GammaV 1 :seq: phin 1  .->.  phin 2
              , GammaV 2 :seq: phin 1
              ] :tf: GammaV 1 :+: GammaV 2 :seq: phin 2

universalInstantiation = [ GammaV 1 :seq: lall "v" (phi 1)]
                        :tf: GammaV 1 :seq: phi 1 tau
\end{lstlisting}
In {\tt modusPonens}, by using functions like {\tt phin} (provided by
generically derived typeclasses) for our various lexical items, we end up with
a rule that can be applied to any language incorporating the lexical module that
provides schematic sentence letters and the lexical module that provides Boolean
connectives. Similarly for {\tt universalInstantiation}.

The generic approach to inference checking makes it possible to rapidly develop
a wide variety of formal systems (since one essentially just has to transcribe
the rules from a textbook). This makes it practical to support systems of logic
from a wide variety of common textbooks, including free and open-source
textbooks which generally do not provide complementary proof-checking software.

\section{Applications}
\label{sec:applications}

In this section, we'll describe some current applications of Carnap. In the
first part, we describe where and how Carnap is currently used, and how it has
been received by students. In the second part, we give more detailed
descriptions of several of the widgets provided by Carnap-GHCJS.

\subsection{Adoption and Reception}
\label{sub:adoption_and_reception}
Carnap has been used successfully in the philosophy department at Kansas State
university to teach a majors-level symbolic logic course (PHILO320), an
introductory course in logic aimed at non-majors (PHILO110), and an advanced
independent study on modal logic (PHILO680). It has also been used and as
a supplementary tool for teaching logic at the University of Birmingham in the
UK, and is currently being used as the central tool for a course in mathematical
logic at the University of North Carolina-Chapel Hill. 

In PHILO320 and PHILO110, students are introduced to Carnap through an online
textbook, \emph{The Carnap Book}~\cite{carnap-book}, provided by Carnap-Server
at Carnap's main website. The chapters of the book are web-pages which
incorporate JavaScript widgets generated by Carnap-GHCJS.  These widgets present
students with interactive exercises that challenge them to apply the concepts
they acquire from the textbook as soon as possible after reading.  The Carnap
Book's interactive exercises offer rapid feedback on work, helping students zero
in on the correct understanding of the ideas presented in the textbook while
these ideas are still fresh. In PHILO680, assignments are provided to students
on problem-sheet web-pages that also incorporate Carnap-GHCJS's widgets.

Both the assignment sheets for PHILO680, and the chapters of the Carnap Book,
are generated by Carnap-Server from text files written in a version of the
pandoc markdown language~\cite{pandoc}, extended with a small amount of extra
syntax used to indicate the insertion of exercises. Using a markdown language
greatly simplifies the process of generating and modifying assignments.

Student feedback at Kansas State has been extremely positive, with the course
consistently receiving scores in the top decile for every category of
evaluation.

\subsection{Widgets and The Carnap Book}

Students interact with The Carnap Book through a range of different widgets
provided by Carnap-GHCJS, which are embedded in web-pages served by
Carnap-Server. At the moment, four main widgets are used in teaching.

\begin{figure}
  \caption{Syntax-Checker Exercise}\label{fig:syncheck}
  \centering\includegraphics[width=\textwidth]{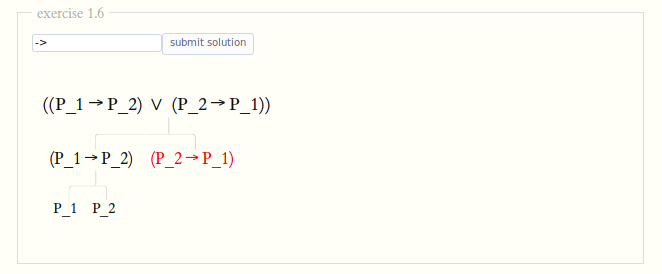}
\end{figure}

The first of these is a ``syntax-checker'' widget, which The Carnap Book uses to
familiarize beginning students with the five standard propositional connectives,
and to help them acquire the concepts \emph{main connective} and \emph{parsing
tree} that they will need for the rest of the course. The task is to enter the
main connective of a highlighted formula into an input field. When the task is
successfully completed, this will ``break down'' the formula into its parts. The
parts are then highlighted and broken down by the student, until the full
parsing tree of the formula is generated. The widget has two variants.  The
first variant (depicted in Figure \ref{fig:syncheck}) asks students to break
down formulas with fully visible parentheses. The second variant familiarizes
them with fixity conventions by asking them to break down formulas with some
implicit parentheses.

\begin{figure}
  \caption{Hardegree-Style Proof (Currying a proposition)}
  \label{fig:Hardegree_proof}
  \centering\includegraphics[width=\textwidth]{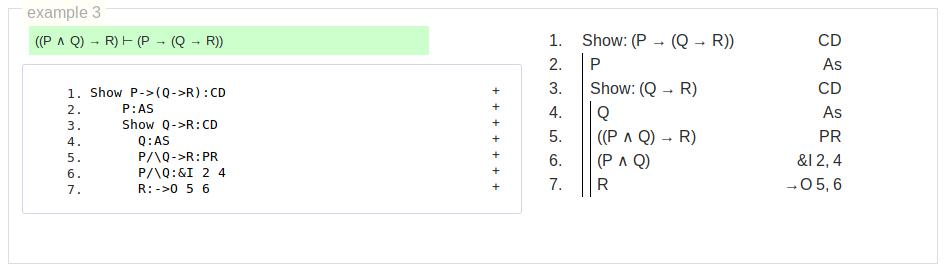}
\end{figure}

\begin{figure}
  \caption{Fitch-Style Proof (Russell's Paradox)}
  \label{fig:fitch_proof}
  \centering\includegraphics[width=\textwidth]{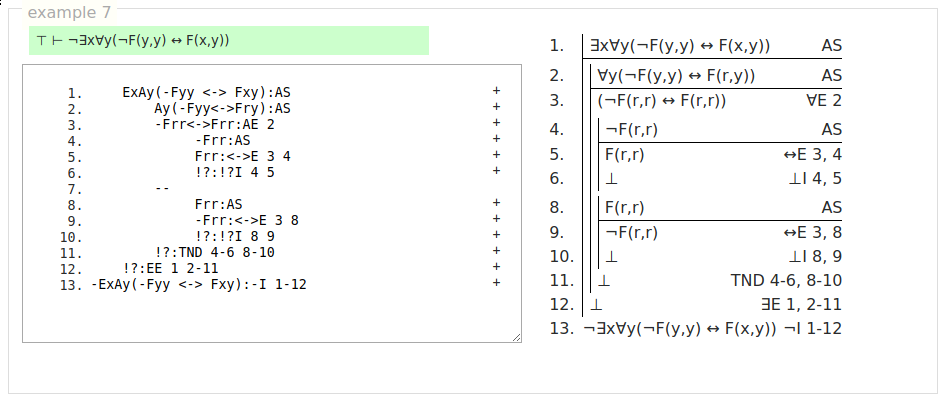}
\end{figure}

\begin{figure}
  \caption{Montegue-Style Proof, in a system of second-order logic}
  \label{fig:montegue_proof}
  \centering\includegraphics[height=7cm]{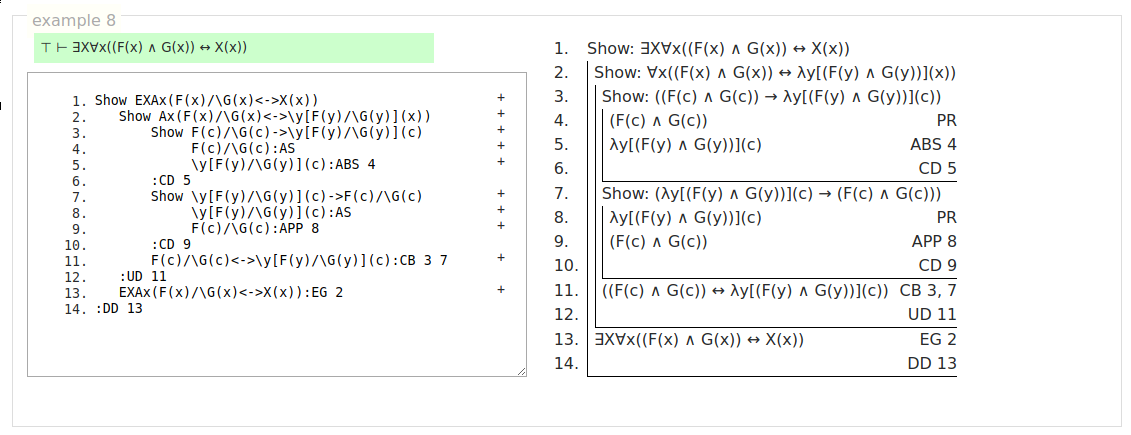}
\end{figure}

Carnap's proof-checking functionality is exposed to students through
a proof-checker widget.  This widget consists of an HTML textarea where students
enter proofs, decorated with a overlay for displaying line numbers and
interactive feedback. In the spirit of flexibility---one of the design
objectives described in Section \ref{sub:design_principles}---the proof-checker
widget allows for three different styles of natural deduction: Hardegree (Figure
\ref{fig:Hardegree_proof}), Fitch (Figure \ref{fig:fitch_proof}), and Montegue
(Figure \ref{fig:montegue_proof}). Feedback on individual proof lines is
available by means of a column, placed to the right of the proof, which displays
a ``+'' for correctly derived lines, and which displays various icons for
different types of error, with more detailed descriptions of problems visible
upon mousing over the icon. Optionally, Carnap can generate a rendering of the
proof closer to what might appear in a textbook, and display the rendering
alongside the student proof (visible in
Figures~~\ref{fig:Hardegree_proof}, \ref{fig:fitch_proof},~\ref{fig:montegue_proof}).
Experimentally, Carnap can also overlay other feedback on the textarea, e.g.
vertical guidelines to help with the proper indentation of subproofs.

\begin{figure}
  \caption{Truth-Table Exercise}\label{fig:truth_tables}
  \centering\includegraphics[width=\textwidth]{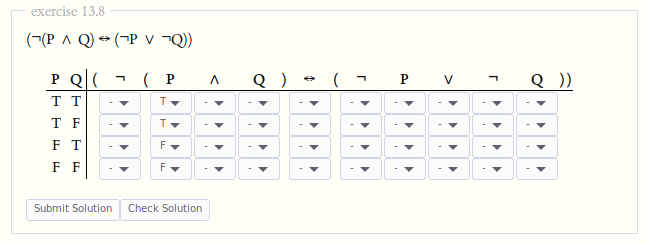}
\end{figure}

To introduce students to semantic methods, The Carnap Book uses a simple truth
table widget (Figure \ref{fig:truth_tables}). Students are able to fill in the
entire truth table to demonstrate the validity of a given argument, or they can
present a single counterexample in order to demonstrate the invalidity of the
argument.  The semantic functions here---checking that a particular sentence is
true relative to a certain assignment of truth values to its sentence
letters---is derived from semantic functions at the level of the lexical modules
that go into the construction of Carnap's propositional language. It would be
possible to use the same method to construct exercises for checking
counterexamples to arguments formulated in the languages of first-order or modal
logic, although this functionality hasn't yet been required.

Finally, with a translation widget, The Carnap Book supplies two forms of
translation exercise, asking students to translate English sentences into the
languages of propositional and first-order logic. In the propositional case,
Carnap can check solutions for correctness up to logical equivalence with the
intended translation. In the first-order case, translations are currently
required to be correct verbatim. Overcoming this limitation will require
introducing a check for equivalence of sentences in the language of first-order
logic, or some fragment thereof---likely through the introduction of an external
theorem-prover. (See the next section for discussion of this possibility.)


\section{Longer Term Prospects}
\label{sec:longer_term_prospects}

In this section, we discuss the longer term prospects of the Carnap
framework. In Subsection~\ref{sub:projected_development}, we discuss new features
we hope to add to the framework, new topics that we hope to incorporate into
educational materials available on the Carnap website, and changes to the
underlying technologies that we may need to address in the future. In
Subsection~\ref{sub:projected_applications}, we discuss some potential future
applications that we envision for the framework.

\subsection{Projected Development}
\label{sub:projected_development}

Carnap's successful use in the classroom and as a platform for the rapid
development of a range of different proof-checking tools suggests that the basic
architecture of the project is sound, and that the fundamental data types
underlying Carnap's abstract syntax are sufficiently flexible to support the
objectives of the project. Ultimately, depending on future development of the
GHCJS compiler, it may make sense to move to the emerging WebAssembly standard,
as it seems likely\footnote{On the basis of e.g.\ the high-level goals of the
WebAssembly team~\cite{webassembly}, and the existence of projects like the
recently funded WebGHC~\cite{webghc}} that Haskell compiled to WebAssembly will
eventually be efficiently executable in the browser. However, major revisions to
the build process or to the fundamental parts of Carnap-Core seem unlikely in
the relative short term. Most near future development will instead focus on
extending Carnap's capacities, and improving the user-facing parts of the
framework.

We are particularly concerned to:
\begin{enumerate}[label=(\alph*)]
  \item implement new unification algorithms;
  \item\label{item:new_languages} continue to extend Carnap's support for a wide
    range of languages and logics; 
  \item develop novel web-based interfaces for proof-construction and semantic reasoning; and
  \item improve the Carnap-Server interface for instructors.
\end{enumerate}
We'll discuss each of these objectives in turn.

First---We are concerned to implement new unification algorithms. While Huet's
algorithm is extremely general and performs reasonably well for the type of
problem that it is applied to in Carnap's proof checking,\footnote{The class of
  unification problems that we use the algorithm for at the moment is restricted
  to higher-order pattern matching (not to be confused with unification of
  higher-order patterns). However, it is worth noting that even a very limited
  subset of this family of unification problems---namely second-order pattern
matching---is NP complete. \cite{baxter} } there are other unification
algorithms which are more efficient. It would be good to have them available in
case efficiency becomes a problem in some future application. A likely target
for a more efficient but still powerful unification algorithm would unification
for higher-order patterns (see~\cite[p1041]{handbook}.) In the near future, we
hope to implement this functionality generically for all of Carnap's languages,
likely along the lines of the implementation described in~\cite{nipkow}.

Our second objective is to extend the range of languages and logics that Carnap
supports. While Carnap's current capacities are sufficient for teaching logic at
the elementary level, there are still various mathematically and philosophically
interesting formalisms for which we still do not have proof systems. In order to
continue to improve Carnap's usefulness as part of a logically rigorous
curriculum in philosophy, we would like to add support for different
constructive systems of logic, for philosophically interesting extensions to the
standard languages of classical logic (e.g.\ definite descriptions, abstraction
operators, truth predicates, multi-modal logics, and so on). And in order to
make Carnap useful for students learning about reasoning in a more traditionally
quantitative context---a standard ``introduction to proof'' class in
a mathematics department, for example---it would also be desirable to add
support for some elementary topics in mathematics, by adding, e.g., a language
for graphs, for elementary algebra, and for elementary number theory or the
theory of some other inductive structure.

Our third objective is to continue to develop novel web-based interfaces for
proof construction and semantic reasoning. Several recent projects show the
potential of the modern web for creating compelling interfaces and
visualizations that enable students to bring visual insight to bear on logical
problems, in something like the way they might bring it to bear on problems in
calculus and geometry.\footnote{Two notable examples are The Incredible Proof
  Machine \cite{incredible}, and The Modal Logic Playground \cite{mlplayground}
} Most of these projects, however, have focused on the interface and left the
underlying logical engine relatively simple. We hope to bring similar interfaces
to Carnap. A first step will be to extend the JSON API mentioned in
Section~\ref{sec:overview_of_the_carnap_framework} to better expose the full
functionality provided by Carnap-GHCJS. This will make it possible to build on
existing open-source work of the kind mentioned above by adapting the interfaces
they provide to plug into Carnap. It will also hopefully smooth the path for
future front-end developers to easily make use of Carnap in similar explorations
of visual reasoning, simplifying such projects by providing a ready-to-go
client-side logic engine for JavaScript-based GUIs.

Our final objective is to improve the Carnap-Server interface for instructors.
At the moment, since Carnap has been used primarily by its developers at Kansas
State University, it isn't possible to remove and modify student accounts or to
set due dates for problem sets from The Carnap Book through the Carnap-Server
web interface. In order to make Carnap-Server more accessible to instructors
outside of the university, it will be necessary to make these classroom
management tasks easy to perform through the Carnap website. The Carnap
website's content-creation functions could also be significantly improved. While
at the moment, it's possible to create custom problem-sheets for one's class,
instructors would benefit from some mechanism for creating other instructional
materials as well: HTML slides with embedded logic widgets,\footnote{Carnap's
  widgets can be embedded in HTML slides for presentations---see the slides
  associated with \cite{slides} for an example---but at the moment the slides
  must be created by hand.
} interactive textbooks and lecture notes similar to The Carnap Book, and
automatically graded exams. Once better content-creation functionality is in
place, it also seems desirable to allow instructors to optionally share their
assignments publicly, or with selected others instructors.

\subsection{Projected Applications}
\label{sub:projected_applications}

In coming decades, as computer-assisted proving moves into the mainstream in
mathematics and computer science, students who wish to prove theorems or to
produce formally verified software will need to be familiar with the idea of
formal proof and with the capacities and limitations of proof-assistants.
Researchers outside of mathematics and computer science departments may also
eventually come to benefit from the higher standards of rigor, clarity, and
precision that computer-assisted proving enforces. It is our hope that Carnap
will eventually be able to serve as a bridge, both from standard introductory
logic and discrete mathematics courses to more advanced computer-assisted logic
and mathematics, and from the study of traditional problems in philosophy to
rigorous research in philosophical logic. In this section, we conclude this
paper by briefly describing those two projected applications.

Our experience with Carnap so far suggests that it is a good on-ramp for
students (including the initially quantitatively anxious students found in
philosophy departments) to learn about computer-assisted proof. Anecdotal
evidence suggests less incremental approaches---for example teaching logic using
the Coq IDE---while effective, presents significant challenges even for third or
fourth year students in computer science~\cite{coq1,lambdata}. This is perhaps
unsurprising. Applications like Coq are not designed with students in mind. Some
of their most impressive features (all-encompassing scope, a tight relation to
the production of formally verified software, a small trusted kernel) may even
be counterproductive to pedagogical applications, where one wants to move in
small steps and initially restrict degrees of freedom to those that are actually
relevant to the concepts students are acquiring.

Hence, one possible application for the Carnap framework in the future will be
the development of specialized ``baby'' proof assistants, designed to help
novice students acquire basic concepts---formal inference, structural induction,
Boolean operations, syntactic types---that will make the notation and operations
of ``full-grown'' proof assistants intelligible, smoothing the learning curve in
more advanced classes. Ideally, this familiarization could be achieved either by
introducing these concepts (in particular, induction) within a second-semester
logic course, or by adapting an introductory discrete mathematics course---where
students often encounter induction and rigorous demonstration for the first
time---both to incorporate an early unit on formal proof and to subsequently use
Carnap as a tool for verifying proofs throughout the semester.

Developing an application for this purpose will require the completion of some
parts of goal~\ref{item:new_languages} from~\ref{sub:projected_development}
above.  In particular, it will be necessary to develop languages and logics that
allow for reasoning by structural induction, and ideally that make use of
notation resembling the notation of existing proof-assistants. It will also be
helpful to create educational materials to designed to support instructors
hoping to use Carnap to ease their students into computer-assisted proving. Much
work remains to be done here, and much remains to be learned about the best
approach to this particular pedagogical challenge. However, given the potential
payoff for students of early exposure to computer-assisted proving, this
application is worth exploring.

The second projected application mentioned above was the development of Carnap
as a tool not just for students of logic but also for researchers in
logic-adjacent areas of philosophy---those studying the foundations of
mathematics, the semantics of natural language, truth predicates, reasoning with
imprecise predicates, and so on. Just as researchers in mathematics can prevent
errors and improve the precise communicability of their results by working in
machine-checkable formalisms, philosophers whose work involves proof stand to
benefit from the use of formal verification technologies.  However, logical work
in philosophy differs from research in pure mathematics in various ways. In
particular, the work is more ``comparative''. In addition to proving theorems,
philosophical logicians tend to be interested in finding formal models for
phenomena of philosophical interest, which often requires looking at the
behavior of a variety of different formalisms or developing entirely new
formalisms.  A framework like Carnap, with its emphasis on flexibility and on
lowering the barrier to entry for users, would be well-suited to philosophical
research. 

Adapting Carnap for this application will again require the completion of some
parts of goal~\ref{item:new_languages} from
Section~\ref{sub:projected_development}. In this case, it will be important to
provide support for formal languages studied in philosophical logic, and to
improve Carnap's interface in a way that makes it suitable for sustained
reasoning and not just solving classroom problems. In particular, it will be
desirable to add formatting tools that will ease the construction and
organization of longer proofs or of reasoning that adopts a theorem/lemma
organization, and to integrate with one or more theorem-provers in order to
shorten some of the more tedious and routine parts of proof-construction. Again,
much work remains to be done, and much remains to be learned about the best
approach to integrating computer-assisted proving into philosophical research.
But the potential benefits---in terms of improved rigour, improved
communicability, and potential reorientation of some forms of research---make
this application worth exploring.

\bibliography{Paper}
\bibliographystyle{eptcs}

\end{document}